\documentclass[12pt]{article}
  
\usepackage{xspace}  
\usepackage{epsfig}  
  
\usepackage{graphicx}               

\newcommand{\nc}{\newcommand}  
  
\nc{\beq}{\begin{equation}}  
\nc{\eeq}{\end{equation}}  
\nc{\beqa}{\begin{eqnarray}}  
\nc{\eeqa}{\end{eqnarray}}  
\nc{\bea}{\begin{eqnarray}}  
\nc{\eea}{\end{eqnarray}}  
\nc{\ra}{\rightarrow}  
\nc{\lsim}{\begin{array}{c}\,\sim\vspace{-21pt}\\< \end{array}}  
\nc{\gsim}{\begin{array}{c}\sim\vspace{-21pt}\\> \end{array}}  
  
\nc{\LL}{L}  
\nc{\vv}{\tilde{v}}  
\nc{\GG}{\widetilde{G}}

\title{  
\vspace*{-2.3cm}  
\begin{flushright}  
\normalsize{  
ANL-HEP-PR-04-88\\
CERN-TH/2004-208\\
CU-TP-1119\\ 
EFI-04-32\\  
FERMILAB-PUB-04/226-T  
  }  
\end{flushright}  
\vspace{1.5cm}  
\Large  
\textbf{Warped Fermions and Precision Tests}\vspace*{1.0cm}  
\author{\large\textbf{Marcela Carena$^a$}, \textbf{Antonio Delgado$^{b,c}$},  
\textbf{Eduardo Pont\'{o}n$^{a,d}$},\\[0.3cm]  
\textbf{Tim M.P. Tait$^{a,e}$},  
and  
\textbf{C.E.M.~Wagner$^{e,f}$}\\ \\[0.5cm]  
$^a$\normalsize\emph{Fermi National Accelerator Laboratory,  
P.O. Box 500, Batavia, IL 60510, USA} \\  
$^b$\normalsize\emph{Department of Physics and Astronomy, Johns
Hopkins University,}\\
\normalsize\emph{3400 North Charles St., Baltimore, MD 21218, USA} \\  
$^c$\normalsize\emph{TH-Division, CERN, 1211 Geneva, Switzerland}\\
$^d$\normalsize\emph{Department of Physics, Columbia University,}\\
\normalsize\emph{538 W. 120th St, New York, NY 10027, USA} \\  
$^e$\normalsize\emph{HEP Division, Argonne National Laboratory,  
9700 Cass Ave.,  
Argonne, IL 60439, USA} \\  
$^f$\normalsize\emph{Enrico Fermi Institute, Univ. of Chicago, 5640  
Ellis Ave., Chicago, IL 60637, USA}}
\date{}}  
\begin{document}  
\setcounter{page}{0}  
\maketitle  
\begin{abstract}  
We analyze the behavior of Standard Model matter propagating in a
slice of AdS$_5$ in the presence of infrared-brane kinetic terms.
Brane kinetic terms are naturally generated through radiative
corrections and can also be present at tree level.  The effect of
the brane kinetic terms is to expell the heavy KK modes from the
infrared-brane, and hence to reduce their coupling to the localized
Higgs field.  In a previous work we showed that sizable gauge kinetic
terms can allow KK mode masses as low as a few TeV, compatible with
present precision measurements.  We study here the effect of fermion
brane kinetic terms and show that they ameliorate the behavior of the
theory for third generation fermions localized away from the infrared
brane, reduce the contribution of the third generation quarks to the
oblique correction parameters and mantain a good fit to the precision
electroweak data for values of the KK masses of the order of the weak
scale.
\end{abstract}  
  
\thispagestyle{empty}  
\newpage  
  
\setcounter{page}{1}

\baselineskip18pt  
  
\section{Introduction}  
\label{sec:intro}  
  
The hierarchy between the apparent Planck scale and electroweak scale
is very mysterious, and leads us to believe that the Standard Model
(SM) is most likely only an effective theory which breaks down at the
electroweak scale.  This belief in fact drives much of the present
activity in particle physics, both to propose alternatives for physics
beyond the SM and to explore their consequences.  One of the more
intriguing possibilities is the Randall-Sundrum (RS)
model~\cite{Randall:1999ee}.  This model invokes a warped metric (of
curvature $k$) to explain how the two scales can coexist quite
naturally: the Planck scale $M_{P}$ is the fundamental scale of the
bulk as a whole, and is the apparent scale for gravity as a result of
the graviton wave function having most of its support at the point
where the warp factor is largest.  The Higgs potential,
however, is naturally at the weak scale as a result of it living on
the other side of the extra dimension where the warp factor renders
the natural scale of order $M_{P} e^{-kL} \sim 1$ TeV. This fixes the
size of the extra dimension such that $kL \sim 30$.
  
The RS hierarchy solution requires only that the Higgs be confined to
the IR boundary at $y=L$ ($y$ is the coordinate in the compact
dimension).  It does not require that the rest of the SM fields be
with the Higgs on the IR boundary.  A particularly attractive
extension has gauge fields and fermions in the bulk
\cite{gaugerunning}-\cite{Huber:2001gw}, allowing one to address grand
unification, absence of TeV-scale FCNC effects, and perhaps even the
observed flavor structure for the SM fermions itself.  The AdS/CFT
connection allows such theories an alternate interpretation as a
nearly conformal 4d theory, with conformal breaking at the TeV scale
\cite{rscft}. 
  
Having promoted the SM gauge fields and fermions to extra-dimensional
fields, the possibility arises that there will be brane interactions
which mimic their 5d kinetic terms
\cite{Georgi:2000ks}-\cite{delAguila:2003kd}.  These terms appear as
irrelevant operators in the 5d theory, but as all interactions are
already irrelevant, they should be considered a generic feature of any
effective theory for extra dimensions.  A specific UV completion could
in principle predict their size, but in the absence of one, they are
part of the most general extra-dimensional model which one may
consider.  Further, as is usually the case with any term not forbidden
by a symmetry, they will be generated radiatively even if the
underlying physics renders them small at tree level.  These terms are
of particular importance on the IR brane of a warped theory, where the
warping enhances their impact, and it is therefore important to study
their physical effects.
  
The theory with gauge fields in the bulk can potentially feel strong
bounds from precision electroweak (EW) data
\cite{Csaki:2002gy}-\cite{Agashe:2004ay}.  The issue is that the
localized Higgs VEV can induce substantial mixing between the ordinary
$W$ and $Z$ bosons with their KK brethren, distorting their properties
at a level in disagreement with precision data.  In the most simple
theories, bounds on the order of 20 TeV can be derived, rendering the
theory impossible to discover at future colliders and re-introducing
fine-tuning in the Higgs potential at a level of $10^{-3}$.

Some of these constraints may be ameliorated by including brane
kinetic terms for the gauge bosons~\cite{Carena:2002dz} 
or by imposing a custodial $SU(2)$
symmetry~\cite{Agashe:2003zs}. 
 Still others could be improved by moving the fermions away
from the IR brane.  However, one quickly runs into a problem with this
second solution: the large top mass indicates that the top is strongly
coupled to the Higgs, and in fact the Kaluza-Klein (KK) modes are even
more strongly coupled than the zero mode.  If the top lies far from
the IR brane, the KK modes become so strongly coupled that the theory
quickly loses a perturbative description.  Thus, there is a kind of
``tug of war'' between the requirements of small EW corrections and a
perturbative top Yukawa interaction.
  
A separate issue related to the geometrical picture of fermion flavor
arises from non-universal couplings which distinguish the bottom
quark.  The small masses of the first and second generation fermions
motivate their being located close to the Planck brane, whereas the
large top mass requires that the top-quark (including left-handed top,
and thus also left-handed bottom) be located close to the TeV brane.
This leads to a non-oblique correction to the
$Z$-$b_L$-$\overline{b}_L$ vertex which ruins the observed agreement
of the predicted $R_b$ with its measured value and can induce
flavor-changing neutral currents at an unacceptable level.
  
In this article we explore a new class of brane kinetic terms, those
relevant for the fermion fields.  As argued above, they are present in
any self-consistent description of the extra dimension anyway, and
they further have great potential to relax some of the EW precision
bounds.  In particular, since they expell the KK modes of the fermions
from the IR brane, they allow for a wider region of localized top
quarks without the strong coupling problems alluded to above.  This in
itself allows one to consider regions of parameters where the
couplings of the SM fermions to the gauge KK modes are strongly
suppressed or may altogether vanish.  As a result, contributions to
the $S$ parameter and possible additional four fermion interactions
can be made small, which can lead to a much more comfortable situation
from the perspective of the EW fit.  They also suppress some
contributions to the $T$ parameter from KK modes of top, and thus are
directly helpful in their own right.  There are still potentially
large contributions to the $T$ parameter coming from the nonuniversal
modification of the $W$ and $Z$ gauge boson wavefunctions, that arises
from the mixing with their KK modes through the localized Higgs.  As
mentioned above, these effects can be suppressed either by including
brane kinetic terms for the gauge fields, or by imposing a custodial
symmetry.  The net result is that in the presence of fermion brane
kinetic terms the EW fit allows lower KK mode masses than in a theory
without the fermion brane kinetic terms, and thus more opportunity to
observe KK modes at future colliders and less EW fine-tuning in the
Higgs potential.
  
This article is organized as follows.  In Section~\ref{sec:kk} we
introduce brane kinetic terms for the even components of bulk
fermions, and derive the spectrum and wave functions.  In
Section~\ref{sec:strong} we examine the role such terms play in strong
coupling limits for fields coupled to the IR brane.  We find that even
for moderate values of the infrared brane kinetic term coefficient,
the constraint from the top quark mass on the fifth dimensional Yukawa
coupling is significantly relaxed.  In Section~\ref{sec:low} we
examine the implications for the EW fit.
Finally, in Section~\ref{sec:conclusion} we conclude.
  
\section{5D Lagrangian and KK Decomposition}  
\label{sec:kk}  
We begin by setting up notation, and deriving the KK decomposition for
a bulk fermion, including brane kinetic terms.  The background metric
can be written as 
\beq
\label{metric}  
ds^2=G_{MN}dx^Mdx^N=e^{-2\sigma}\eta_{\mu\nu}dx^\mu dx^\nu+dy^2~,  
\eeq  
with $\eta_{\mu\nu}=\mbox{diag} (-1,+1,+1,+1)$, $\sigma(y)=k|y|$ and  
$0\leq y\leq L$.    
We use upper case roman letters for the 5d Lorentz indices, and lower
case greek letters for the 4d ones.  The Higgs field is localized on
the $y=L$ brane (IR brane) where the fundamental scale is red-shifted
to TeV values, thus solving the hierarchy problem.  We assume that
both the standard model gauge bosons and fermions live in the bulk,
together with gravity.
  
The lagrangian for a freely gravitating fermion, including brane
localized kinetic terms, can be written as
\beq  
S=-\int d^4x \, \int_0^L dy \, \sqrt{-G}\, \{i\bar{\Psi}
\Gamma^A e_A^M D_M \Psi+  
i M(y) \bar{\Psi}\Psi  
+2 \alpha_{f}\delta(y-L) i\bar{\Psi}_L \gamma^a e_a^\mu 
\partial_\mu \Psi_L\}~.  
\label{action}  
\eeq  
\noindent  
We use ($\Gamma$, $\gamma$) for (5d,4d) $\gamma$-matrices\footnote{For
definiteness, we use the following representation of the 5-D
$\Gamma$-matrices:
\beqa
\begin{array}{cc}
\Gamma^\mu = \left( \begin{array}{cc}
0 & \sigma^\mu \\ \bar{\sigma}^\mu & 0
\end{array} \right) &
\hspace{1cm}
\Gamma^5 = \left( \begin{array}{cc}
1 & 0 \\ 0 & -1
\end{array} \right)
\end{array}~,
\nonumber
\eeqa
where $\sigma^\mu = (1, \vec{\sigma})$, $\bar{\sigma}^\mu = (-1,
\vec{\sigma})$ and $\vec{\sigma}$ are the Pauli matrices.  We also
define the chirality projectors by $P_{L(R)} = \frac{1}{2} (1 \pm
\Gamma^5)$.} and ($A$, $a$) for the (5d,4d) tangent-space Lorentz
indices.  $G$ represents the determinant of the (5d) metric, $e$ is
the vielbein, $D_M$ is the covariant derivative, including the spin
connection, and $\alpha_{f}$ is the coefficient of the brane localized
kinetic term.  Note that $\alpha_{f}$ has dimension of ${\rm
mass}^{-1}$.  The $\delta$-function is normalized so that
$\int_{0}^{\LL} 2 \delta(y) dy = 1$.
  
The boundary conditions at $y=0,L$ are chosen so that the low-energy
theory is chiral \cite{Gherghetta:2000qt}.  For definiteness, in the
above case only the left-handed component $\Psi_L$ has a zero mode.
The mass function is $M(y) =c_{f} \sigma^\prime$ 
(i.e., it is an odd mass term), where the dimensionless
bulk mass parameter $c_{f}$ essentially determines the localization of
the massless (zero) mode.
  
The above action is not the most general one at the quadratic level.
We are only including brane terms on the IR brane, and then only so
for the even chirality and involving $\partial_\mu$ as opposed to
$\partial_5$.  
The first choice is purely for phenomenological purposes, 
since the UV brane kinetic terms are irrelevant for the KK mode spectrum.
One way
to understand this is that the wave function of KK modes whose masses
are $\sim O({\rm TeV})$ are localized near the IR brane and are
therefore relatively insensitive to the UV brane terms.
We will briefly consider 
the effect of UV brane terms on the EW fit in Sec.~\ref{sec:gauge} below.  

The second choice, adding kinetic terms only for the even field
components, can be thought of as a prescription for some of the UV
physics.  In the absence of localized kinetic terms, the even fields
($\Psi_L$) will couple to the brane whereas the odd fields ($\Psi_R$)
will not (the odd wave functions vanish on the brane as a result of
the odd boundary conditions), and therefore operators like
$i\bar{\Psi}_R \gamma^a e_a^\mu \partial_\mu \Psi_R$ vanish on the
brane.
Furthermore, if this term is absent, it will not be
perturbatively generated, and thus this situation is technically
natural.  

One may still consider the nonvanishing operator $i(\bar{\Psi}_L
\partial_{5} \Psi_R + \partial_{5} \bar{\Psi}_R \Psi_{L})$ localized
on the brane, although its interpretation requires a careful
regularization of the brane thickness.  
From a
practical point of view, the choice of Eq.~(\ref{action}) is
convenient because it insures that the $\Psi_L$ wave functions are
continuous on the brane, and thus their couplings to brane fields are
well-defined in the infinitely narrow brane approximation.

\subsection{KK Decomposition}
\label{sec:kkdecomp}
  
We expand the fermion field in KK modes as  
\bea  
\Psi_{L,R}(x,y)=e^{3\sigma/2} \sum_n \psi^n_{L,R}(x)f_{L,R}^n(y)~,  
\eea 
where the KK mode wavefunctions, $f^n_{L,R}$, satisfy the set of
coupled equations
\bea  
\left[\partial_5 + (c_{f}-1/2)\sigma'\right] f_L^n&=&e^\sigma m_n f_R^n~,  
\label{eqn1} \\  
\left[-\partial_5 + (c_{f}+1/2)\sigma'\right] f_R^n&=&e^\sigma m_n f_L^n   
\left[1+2 \alpha_{f}\delta(y-L)\right]~,  
\label{eqn2}  
\eea  
and $m_{n}$ are the KK masses.  In order to have canonically
normalized kinetic terms in the 4d KK description, we choose the
wavefunctions $f_{L,R}$ to satisfy the following orthonormality
relations
\bea  
\int_0^Ldy\left[1+2\alpha_{f}\delta(y-L)\right] f^n_ {L} f^m_{L} &=& 
\delta_{mn}  
\nonumber \\  
\int_0^Ldy  f^n_ {R} f^m_{R} &=& \delta_{mn} ~.  
\label{normalization}  
\eea  
  
The appropriately normalized zero-mode wavefunction is 
\bea  
f_L^0(y)&=&\sqrt{ \frac{k(1-2c_{f})}  
{e^{(1-2c_{f})kL}
\left[ 1+(1-2c_{f})\alpha_{f} k \right]-1} } 
e^{(1/2-c_{f})\sigma}~,  
\label{zeromodefermion}
\eea  
and the odd tower, by construction, does not contain a zero-mode.  To
solve for the massive KK mode wavefunctions we use Eq.~(\ref{eqn1}) to
find $f_R^n$,
\beq  
\label{solfR}  
f_R^n = \frac{e^{-\sigma}}{m_{n}} \left[ \partial_{5} +   
(c_{f} - 1/2) \sigma' \right] f_{L}^{n}~,  
\eeq  
and replacing in Eq.~(\ref{eqn2}) we have a second order differential
equation for $f_L^n$:
\beq  
\left[-\partial_{5}^2 + 2\sigma' \partial_{5} - \left(1 -   
\left( c_{f} + 1/2 \right)^2 \right)(\sigma')^2   
- (c_{f}-1/2)\sigma{''} \right] f_L^n=e^{2\sigma}m_n^2   
\left[1+2\alpha_{f}\delta(y-L)\right] f_L^n  
\label{eqfL}~.  
\eeq  
Using $\sigma'' = 2 k \left[ \delta(y) - \delta(y-L) \right]$, we see  
that we have the boundary conditions  
\beqa  
\left. \rule{0mm}{5mm} \partial_{5} f_{L}^{n} \right|_{y=0} &=& \left.   
\left( \frac{1}{2} - c_{f} \right) k f_{L}^{n}  \right|_{y=0}~,  
\label{condition0} \\  
\left. \rule{0mm}{5mm} \partial_{5} f_{L}^{n} \right|_{y=L} &=& \left[  
\left( \frac{1}{2} - c_{f} \right) k  +   
\left. \rule{0mm}{5mm} \alpha_{f} m_{n}^{2} e^{2kL} \right]  
f_{L}^{n}  \right|_{y=L}~.  
\label{conditionL}   
\eeqa  
The solution to Eq.~(\ref{eqfL}) in the bulk is  
\beq  
f^n_L(y)=A_n e^\sigma \left[ J_{|c_{f}+1/2|}\left(\frac{m_n}{k}
e^\sigma\right)+b_{n}   
J_{-|c_{f}+1/2|}\left(\frac{m_n}{k}e^\sigma\right)\right]~,  
\label{wave}  
\eeq  
where $A_{n}$ is fixed by the normalization condition
Eq.~(\ref{normalization}), while the boundary conditions,
Eqs.~(\ref{condition0}) and (\ref{conditionL}), determine $b_{n}$ and
the KK spectrum by
\bea  
b_{n}&=&-\frac{(c_{f}+1/2)J_{|c_{f}+1/2|}\left(\frac{m_n}{k}\right)+  
\frac{m_n}{k}J'_{|c_{f}+1/2|}\left(\frac{m_n}{k}\right)}{(c_{f}+1/2)
J_{-|c_{f}+1/2|}  
\left(\frac{m_n}{k}\right)+  
\frac{m_n}{k}J'_{-|c_{f}+1/2|}\left(\frac{m_n}{k}\right)}\nonumber\\  
&=&-\frac{(c_{f}+1/2-\tilde{\alpha}_{f}\frac{m_n^2}{\tilde{k}})J_{|c_{f}+1/2|}  
\left(\frac{m_n}{\tilde{k}}\right)+  
\frac{m_n}{\tilde{k}}J'_{|c_{f}+1/2|}\left(\frac{m_n}{\tilde{k}}\right)}  
{(c_{f}+1/2-\tilde{\alpha}_{f}\frac{m_n^2}{\tilde{k}})J_{-|c_{f}+1/2|}  
\left(\frac{m_n}{\tilde{k}}\right)+  
\frac{m_n}{\tilde{k}}J'_{-|c_{f}+1/2|}\left(\frac{m_n}{\tilde{k}}\right)}~,  
\label{KK}  
\eea 
where we defined $\tilde{k}=k e^{-kL}\sim O({\rm TeV})$ and
$\tilde{\alpha}_{f}=\alpha_{f} e^{kL} \sim O({\rm TeV}^{-1})$.  Note
that like any other IR brane term, the localized fermion kinetic term
is warped, and as a term with inverse mass dimension, the warping
increases its importance at low energies.  As usual, the consistency
of the boundary conditions is what determines the mass eigenvalues
$m_n$.
  
We can obtain approximate expressions for the wavefunctions of the  
lowest lying KK modes ($m_n \ll k$).  When $c_{f} > \frac{1}{2}(1+1/kL)$,  
the eigenvalue equation reduces to $b_{n} = {\cal{O}}(m_{n}/k)^{2c_{f}-1}  
\ll 1$, i.e.  
\beq  
J_{c_{f}-1/2}(x_n) \approx \alpha_{f} k x_nJ_{c_{f}+1/2}(x_n)~,  
\eeq  
where $m_n \equiv x_n \tilde{k}$, and the wavefunctions become  
\bea  
f^n_L(y)&\approx&A_n e^{ky} J_{c_{f}+1/2}(x_n e^{-k(L-y)})~,\nonumber\\  
A_n&=&\frac{e^{-kL}}{J_{c_{f}+1/2}(x_n)}\sqrt{\frac{2k}{1+(1-2c_{f}) 
\alpha_{f} k + \alpha_{f}^2 k^{2} x_n^2}}~.
\label{normalizedf1}  
\eea  
In this case, when the localized kinetic term is large, $\alpha_{f} k
\gg 1$, one of the modes becomes light\footnote{However, in the large
$\alpha_{f}$ limit and making no approximations, the mass never goes
to zero, but asymptotes to the value
$\sqrt{4c_{f}^{2}-1}\,e^{-(c_{f}-1/2)kL}\tilde{k}$.} with $m \simeq
\sqrt{(2c_{f}+1)/(\alpha_{f} k)}\,\tilde{k}$.
  
When $c_{f} < \frac{1}{2}(1-1/kL)$, the eigenvalue equation reduces to  
\beq  
J_{1/2-c_{f}}(x_n) \approx - \alpha_{f} k x_nJ_{-c_{f}-1/2}(x_n)~,  
\eeq  
and the wavefunctions are now given by
\bea  
f^n_L(y)&\approx&A_n e^{ky} J_{-c_{f}-1/2}(x_n e^{-k(L-y)})~,\nonumber\\  
A_n&=&\frac{e^{-kL}}{J_{-c_{f}-1/2}(x_n)}\sqrt{\frac{2k}{1+(1-2c_{f}) 
\alpha_{f} k + \alpha_{f}^2 k^{2} x_n^2}}~.
\label{normalizedf2}  
\eea  
In this case there is no light mode.  

For $c_{f}=1/2$, the wavefunctions read 
\beq
f^n_L(y)=A_n e^\sigma \left[ J_{1}\left(\frac{m_n}{k}e^\sigma\right)+b_{n} 
Y_{1}\left(\frac{m_n}{k}e^\sigma\right)\right]~,
\label{fconefalf}
\eeq
and the eigenvalues are now given by
\bea
b_{n} &=&  - \frac{J_0 \left( \frac{m_{n}}{k} \right)}
{Y_0 \left( \frac{m_{n}}{k} \right)}  \nonumber \\
&=&  - \frac{J_0 \left( \frac{m_{n}}{k} e^{k L}  \right)
		 - m_{n} \, \tilde{\alpha}_{f}
J_1 \left( \frac{m_{n}}{k} e^{k L} \right)}
{Y_0 \left( \frac{m_{n}}{k} e^{k L} \right)
- m_{n} \,\tilde{\alpha}_{f} Y_1 \left( \frac{m_{n}}{k} e^{k L} \right) }~.
\label{eigenmassonehalf}
\eea
The normalization factor, $A_{n}$, is not particularly simple 
and should be calculated from Eq.~(\ref{normalization}).

\subsection{Mixed Position/Momentum Space Propagator}
\label{sec:prop}

When computing loops or performing sums over all KK modes in the
tower, the explicit decomposition is not the most convenient way to
proceed.  It is simpler to employ the propagator in mixed
position/momentum space which implicitly includes the sum over all of
the KK modes.  Thus, in this section we compute the fermion propagator
which shall be of use in Section~\ref{sec:strong} in estimating how
strong a brane coupling involving the fermion can be made before the
theory loses predictivity.

To be definite, we calculate the propagator by assuming that the
boundary conditions are such that the zero mode is left-handed.  The
defining equation for the fermion propagator in the presence of a
brane localized kinetic term for the left-handed components, as in
Eq.~(\ref{action}), is then
\beq  
\label{eqnpropagator}  
i \left( \Gamma^A {e_{\! A}}^{\! M} D_M + M \right) G(X,X') +   
2 \delta(L-y) \alpha_{f} i \Gamma^a {e_{\! a}}^{\! \mu} P_{L} \partial_\mu    
G(X,X') = \frac{i}{\sqrt{-G}} \delta^{(5)}(X-X')~,  
\eeq  
where $M = c_{f} \sigma'$ is the $Z_{2}$-odd bulk mass that determines
the localization of the zero-mode, and $X=(x^{\mu}, y)$.  We Fourier
transform Eq.~(\ref{eqnpropagator}) along the four noncompact
coordinates\footnote{Explicitly, we define $G^{p}(y,y') = \int \!
d^{4}x \, e^{i \eta_{\mu\nu} p^{\mu} x^{\nu}} G(x,y,y')$, so that $p$
is the momentum measured by UV observers.  The $y$-dependent cutoff of
the theory is then $e^{-ky} \Lambda$, where $\Lambda \sim M_{P}$, the
Planck scale.} and define the propagators for the various chiralities
by $G_{LL} \equiv P_{L} G P_{R} = \langle \Psi_{L} \overline{\Psi_{L}}
\rangle$, $G_{RL} \equiv P_{R} G P_{R} = \langle \Psi_{R}
\overline{\Psi_{L}} \rangle$, etc., where $\Psi$ is a generic 5D
fermion and $P_{L,R}$ are the left- and right-handed chirality
projectors.
  
Since only the KK tower that contains a zero-mode couples to the  
brane, we concentrate on the propagator for the left handed  
components, $G_{LL}$.  We may isolate it by first projecting
onto Eq.~(\ref{eqnpropagator}) by $P_L$ from the left and by $P_{R}$  
from the right to obtain  
\beq  
\label{GRL}  
G_{RL} = -i \frac{e^{-\sigma}}{\not{\!p}} \left[ \partial_{y} -   
(2-c_{f}) \sigma'  \right] G_{LL}~.  
\eeq  
Repeating the same projection after applying the operator $i\left(  
\Gamma^A {e_{\!  A}}^{\!  M} D_M - M \right)$ to  
Eq.~(\ref{eqnpropagator}) gives a second equation that relates  
$G_{LL}$ and $G_{RL}$, and using Eq.~(\ref{GRL}) to eliminate $G_{RL}$  
we obtain a second order differential equation for $G_{LL}$:  
\beq  
\left\{ \partial_{y}^{2} - \sigma' \partial_{y} -c_{f}(c_{f}+1) {\sigma'}^{2} 
+ c_{f} \sigma'' - e^{2\sigma} p^{2} \left[ 1 + 2 \delta(L-y) \alpha_{f} \right] 
\right\} G^{p}_{+}(y,y') = - e^{3\sigma} \delta(y-y')~,  
\eeq  
where we defined $G_{+}$ by $G_{LL} = i P_{L} \hspace{-1.5mm}
\not{\!p}\, e^{2\sigma} G_{+}$, and the subscript ``$+$'' refers to
our choice of $Z_{2}$-even boundary conditions for the left-handed
spinor components.  Note that due to the metric signature in
Eq.~(\ref{metric}), $p^{2}<0$ in the on-shell region and the solution
to Eq.~(\ref{eqnpropagator}) can be directly interpreted as the
Euclidean space propagator.  The boundary conditions to be applied on
$G_{+}$ are
\beqa  
\left. \rule{0mm}{4mm} \partial_{5} G_{+} \right|_{y=0} &=&   
\left. \rule{0mm}{4mm} - c_{f} k G_{+}  \right|_{y=0}~,  
\label{conditionProp0} \nonumber \\  
\left. \rule{0mm}{4mm} \partial_{5}G_{+} \right|_{y=L} &=&   
\left. \rule{0mm}{4mm} - c_{f} k G_{+}  \right|_{y=L} -   
\left. \rule{0mm}{4mm} \alpha_{f} p^{2} e^{2kL} G_{+}  \right|_{y=L}~,  
\label{conditionPropL}   
\eeqa  
and the explicit solution is  
\beqa  
\label{propagator}  
e^{2\sigma(y)} G_+^{p}(y,y') &=& - \frac{e^{\frac{5}{2}k(y+y')}}{k(AD - BC)}  
\left[ A K_{c_{f}+1/2} \left( \frac{p}{k} e^{k y_<} \right) +  
B I_{c_{f}+1/2} \left( \frac{p}{k} e^{k y_<} \right) \right] \times
\nonumber \\  
& & \hspace{3cm}  
\left[ C K_{c_{f}+1/2} \left( \frac{p}{k} e^{k y_>} \right) +  
D I_{c_{f}+1/2} \left( \frac{p}{k} e^{k y_>} \right) \right]~,  
\eeqa  
where $K_{\beta}$, $I_{\beta}$ are modified Bessel functions of order  
$\beta$, $y_{<(>)}$ are the smallest (largest) of $y$, $y'$ and  
\beqa  
\label{coefficients}  
A &=& I_{c_{f}-1/2}\left(\frac{p}{k}\right) \hspace{1.2cm}  
C = I_{c_{f}-1/2} \left(\frac{p}{k}e^{k \LL}\right) +  
p  \, e^{k \LL} \alpha_{f} I_{c_{f}+1/2} \left(\frac{p}{k}e^{k \LL}\right) 
\nonumber \\  
B &=& K_{c_{f}-1/2} \left(\frac{p}{k}\right) \hspace{1cm}  
D = K_{c_{f}-1/2}\left(\frac{p}{k}e^{k \LL}\right) -  
p  \, e^{k \LL} \alpha_{f} K_{c_{f}+1/2} \left(\frac{p}{k}e^{k \LL}\right)~.   
\eeqa  
We note that if the boundary conditions are such that the zero-mode is  
right-handed, the propagator for the even components is given by  
$G_{RR} = i P_{R} \hspace{-1mm} \not{\!p}\, e^{2\sigma} G_{+}$, with  
$G_{+}$ given by Eqs.~(\ref{propagator}) and (\ref{coefficients}).  
  
\section{Strong Coupling Estimates}  
\label{sec:strong}  
  
It is interesting to consider the effect of the IR localized kinetic
term on higher order corrections to various observables.  In
particular, since the heavy KK modes are localized close to the IR
brane, corrections that involve brane localized couplings can be quite
important.  In fact, in the absence of localized kinetic terms one
generally encounters enhancement factors $\sqrt{2 k L}$ associated
with every localized coupling that involves heavy KK modes.  This,
combined with sums over the modes inside loops which often diverge,
may cast doubt on the applicability of a perturbative analysis.  The
main effect of the localized kinetic term is to repel the heavy mode
wavefunctions from the brane and therefore one might expect that the
strong coupling effects associated with the KK modes will be
alleviated.
  
An associated point which is relevant for the phenomenology of the
scenario we are considering has to do with the localization of the
standard model fermion zero modes in the extra dimension, i.e. the
choice of the $c$-parameters.  In particular, without the brane
localized kinetic terms the top wavefunctions need to be localized
close to the IR brane ($c_{f} \ll 1/2$) to reproduce the large top
mass without having to introduce too large a 5D Yukawa coupling.  The
presence of a top brane localized kinetic term and the associated
softening of its KK tower couplings can relax such a constraint, and
have an impact on the bounds derived from electroweak precision
measurements.
  
\begin{figure}[t]  
\centerline{\includegraphics[width=0.8\textwidth]{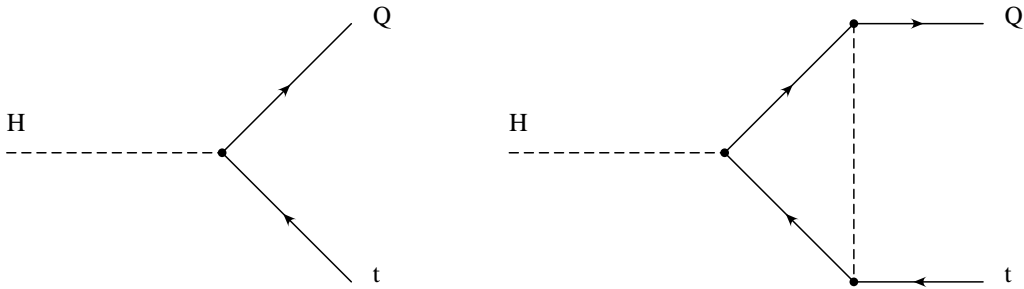}}
\caption{Tree level and one-loop contributions to the 5-dimensional  
top Yukawa coupling, $\lambda_{5}$.  At strong coupling the two  
contributions are comparable.}  
\label{fig:twopoint}  
\end{figure}  
  
With this motivation in mind, we turn to estimate the strong coupling  
bounds coming from the higher dimensional theory on Yukawa couplings,  
e.g.  
\beq  
\int d^4x \int_0^L dy \sqrt{-G} 2 \delta(L-y) \lambda_{5} H \bar{Q} t~.  
\eeq   
To define the strong coupling value of $\lambda_{5}$ we require that  
all loops involving this coupling contribute equally to observables.  The 
presence of both the nontrivial warping and brane kinetic terms change 
the NDA estimate of \cite{Chacko:1999hg}. 
We may estimate this value by calculating the one loop contribution to  
$\lambda_{5}$ itself and requiring that it be as large as the  
tree-level value (see Fig.~\ref{fig:twopoint}).  
The one loop vertex correction to  
$\lambda_{5}$ involves a summation over the KK modes defined in the  
previous section.  An efficient way to sum the KK contributions, which  
also renders the physics more transparent, is to calculate the loop  
directly in the five-dimensional theory using the mixed  
position/momentum space propagator presented in Sec.~\ref{sec:prop}.  
  
Omitting the external propagators and working at zero external momentum,
the loop diagram of Fig.~\ref{fig:twopoint} is,
\beq  
\label{vertexcorrection}  
\left( \lambda_{5} e^{-4kL} \right)^{3} \int \frac{d^{4}p}{(2\pi)^{4}}   
\frac{e^{2kL}}{p^{2}} G_{t_{R}t_{R}}^{p}(L,L) G_{Q_{L}Q_{L}}^{p}(L,L)~,  
\eeq  
where the Higgs propagator contains a factor of $e^{2kL}$ due to not  
being canonically normalized.  We note that for $p\gg k e^{-kL}$,  
\beq  
e^{2kL} G_+^{p}(L,L) \approx \frac{e^{4kL}}{p+ e^{kL} \alpha_{f} p^{2}}~,  
\eeq  
and as a result the $p$ integration in Eq.~(\ref{vertexcorrection}) is  
logarithmically divergent just
as it would have been in a four dimensional theory.  
We can easily estimate the dominant contribution  
by cutting the integration off at $\tilde{\Lambda} \equiv e^{-kL}  
\Lambda$, where $\Lambda \sim {\cal{O}}(M_{P})$,  
(and assuming $\alpha_Q = \alpha_t = \alpha$),
\beqa  
& & e^{-10kL} P_{R} \frac{\lambda_{5}^{3}}{8\pi^{2}} 
\int^{\tilde{\Lambda}}_{\tilde{k}} p^{3} dp   
\left[ \frac{e^{4kL}}{p+ e^{kL} \alpha p^{2}} \right]  
\left[ \frac{e^{4kL}}{p+ e^{kL} \alpha p^{2}} \right]
\nonumber \\  
& =  & 
e^{-4kL} P_{R} \frac{\lambda_{5}^{3}}{8\pi^{2}}   
\frac{1}{\alpha^{2}}  
\left\{ \log\left[ \frac{1+ e^{kL} \alpha \tilde{\Lambda}}{1+\alpha k}   
\right]   
- \frac{e^{kL} \alpha \left(\tilde{\Lambda} - \tilde{k} \right)
}{(1+e^{kL}\alpha \tilde{\Lambda})
(1+ \alpha k)} \right\}~.
\eeqa  
For our estimates we assume for simplicity that the  
$Q_{L}$ and $t_{R}$ brane kinetic terms are of the same order.  
We obtain that the above one-loop contribution 
is smaller than the tree level piece, $e^{-4kL} \lambda_{5}$, when
\beq  
{\lambda}_{5} \lsim 
\frac{\sqrt{8}\pi {\alpha}}{\sqrt{\log [{\Lambda}/{k}]}}~,  
\eeq  
which is a good approximation for $k \alpha \gsim$ a few.
  
This result has interesting implications for the localization of the
top wavefunctions in the extra dimension.  The effective
four-dimensional top Yukawa coupling is
\beq  
\lambda_t = a_{Q} a_{t} \frac{{\lambda}_5}{ \LL}~,  
\label{ytop}
\eeq  
where the parameters  
\beq  
\label{af}  
a_{f} = \sqrt{\frac{(1 - 2 c_{f}) k \LL e^{(1-2c_f)k\LL} }{e^{(1-2c_{f})k \LL}
[1 + (1-2c_f) \alpha_{f} k]-1}}\,  
\approx  
\left \{  
\begin{array}{ll}  
\sqrt{(2 c_{f} - 1) k \LL}\,e^{-(c_f-1/2)k\LL} & \vspace{4.5mm}  
\hspace{0.5cm} \;\; c_f - 1/2 \gsim 1/2k \LL \\  
\sqrt{ \frac{\LL}{\LL + \alpha_{f}}} &  \vspace{3mm} \hspace{0.5cm}
\;\; c_f = 1/2 \\  
\sqrt{\frac{(1-2 c_{f}) k \LL}{1 + (1-2c_f) \alpha_{f} k}}  
 & \hspace{0.5cm}\;\; 1/2 - c_f \gsim 1/2  k \LL
\end{array}  
\right.  
\eeq  
are determined by the localization of the zero mode.  For later  
application we note that one may take $c_{Q} = 1/2$, $c_{t} = 0$,
without reaching the strong coupling regime, provided $\alpha k$ satisfies
\bea
\lambda_t & < &
\frac{\sqrt{8} \pi k \alpha}{\sqrt{\log \left[{\Lambda}/{k}\right]}
\sqrt{k(L+\alpha)(1+k\alpha)}}  ~.
\eea
For $\lambda_t \sim 1$, this is indeed satisfied for
${\alpha} k \gsim$ a few.
  

\section{Low-energy implications}  
\label{sec:low}  
We now consider the effect of fermion brane localized kinetic terms on
the EW fit in the Randall-Sundrum scenario with gauge and fermion
fields in the bulk.  We also include moderate IR localized gauge
kinetic terms since they can have an important impact on the bounds on
the KK spectrum of this class of theories, as was shown in
\cite{CDPTW}.

Another important source of model dependence is related to the
localization in the extra dimension of the fermion zero modes, to be
identified with the SM fields, which is controlled by $c_{f}$.  Since
the KK modes of the gauge fields tend to be localized towards the IR
brane, the couplings of the SM fermions to the gauge field KK modes
depend strongly on the corresponding values of $c_f$.  This implies
that, for KK masses of order a few TeV, we should either have $c_f
\gsim 1/2$ (where such couplings become largely insensitive to the
precise value of $c_{f}$), or choose similar values of $c_f \lsim
1/2$, in order to avoid dangerous flavor changing neutral current
effects.

An attractive idea is that the actual quark and lepton mass
hierarchies, as well as the observed mixing angles, are a consequence
of the fermion localization in the extra dimension.  In such a
scenario, the first two generations are localized closer to the UV
brane ($c_f \gsim 1/2$) to account for the smallness of their masses
compared to the electroweak scale.  The third generation, however,
requires $c_f \lsim 1/2$ to account for the large top mass.  An
important result from the previous section is that in the presence of
moderate localized kinetic terms for the top system, it is possible to
have $c_{t_{L}} = 1/2$, while the right-handed top is localized closer
to the IR brane (with $c_{t_{R}} \sim 0$), without encountering strong
coupling effects due to their KK towers.  Thus, an attractive scenario
emerges where all the fermion fields have $c_f \approx 1/2$ (except
for $t_{R}$) and the quark and lepton mass hierarchies are understood
geometrically.  Here we concentrate on the EW constraints on such a
scenario, since the constraints on models with fermions localized
close to the IR brane (that do not explain the fermion mass
hierarchies) have been explored elsewere\footnote{In this case, the
presence of fermion brane kinetic terms plays an important role in
suppressing the contribution to $T$ from KK top loops, so that it can
be safely neglected, as was argued in \cite{CDPTW}.}
\cite{Carena:2002dz,CDPTW}.  As a first approximation, we consider the
EW fit when all fermions have $c_f = 1/2$ (except for $t_{R}$) and 
similar brane localized kinetic terms so that
the main effect of the new physics is well approximated by the oblique
parameters \cite{Peskin:1991sw} $S$, $T$, and $U$.  In the more
realistic scenario discussed above, one should consider the additional
bounds coming from the flavor nonuniversality, but such effects should
be small for $c_f \gsim 1/2$, and the complete analysis is beyond the
scope of this work.

\subsection{KK Fermion Contributions}  
\label{sec:fermion}  
\begin{figure}[t]  
\centerline{\includegraphics[width=0.4\textwidth]{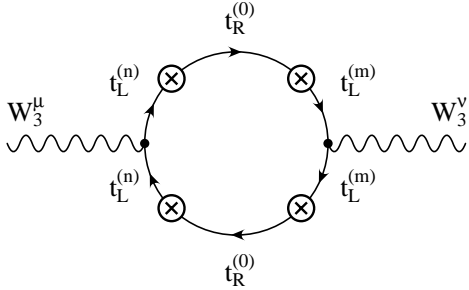}}
\caption{Representative lowest order contributions to the $T$ 
parameter from fermionic KK loop
diagrams.  The crosses represent insertions of the Higgs VEV.  The
corresponding $W^+$-$W^-$ graphs are neglected in the limit $m_t \gg m_b$.}
\label{fig:loop}  
\end{figure}

We start by considering the low-energy consequences from integrating
out the fermion KK modes.  These are loop-level effects, that can
nevertheless be important when the KK fermions couple significantly to
the Higgs.  The most important effect is a contribution to the $\rho$
parameter from KK top loops.  Treating the Higgs VEV perturbatively,
the lowest order contributions arise from diagrams such as those shown
in Fig.~\ref{fig:loop}.  In the absence of fermion brane kinetic
terms, the localized Higgs couplings, which induce mixing among the KK
modes, are independent of the KK level.  As a result the sum over the
KK towers lead to logarithmic and quadratic divergences for the two
diagrams in the figure, respectively.

In the presence of brane kinetic terms, both diagrams become finite
due to the decoupling of the heavier KK modes.  From
Eqs.~(\ref{normalizedf1}) and (\ref{normalizedf2}) we see that, for
$c_{f} \neq 1/2$, for example, the couplings are given by
\beq
\label{KK0couplings}
\lambda_{t_{L}^{(n)},t_{R}^{(0)}} \equiv \lambda_{n0}
= \sqrt{\frac{2k L}{1+(1-2c_{Q}) \alpha_{Q} k 
+ \alpha_{Q}^2 k^{2} x_{Q^{n}}^2}} \, a_{t} \frac{\lambda_{5}}{L}~,
\eeq
when a single KK mode and a zero mode are involved [$a_{t}$ is defined
in Eq.~(\ref{af})], and
\beq
\label{KKcouplings}
\lambda_{t_{L}^{(n)},t_{R}^{(m)}} \equiv \lambda_{nm} = 
\sqrt{\frac{2k L}{1+(1-2c_{Q}) \alpha_{Q} k +
\alpha_{Q}^2 k^{2} x_{Q^{n}}^2}} \sqrt{\frac{2k L}{1+(1-2c_{t}) \alpha_{t} k +
\alpha_{t}^2 k^{2} x_{t^{m}}^2}} \, \frac{\lambda_{5}}{L}~,
\eeq
for the couplings among KK modes.  Imposing that the the top mass be
reproduced determines $\lambda_{5}/L$ from Eq.~(\ref{ytop}).  We see
that indeed the heavier KK modes couple more weakly to the brane.
When $c_{f}=1/2$ one should use the general expressions for the KK
mode wavefunctions, Eq.~(\ref{fconefalf}), although the approximate
expressions, Eqs.~(\ref{KK0couplings}) and (\ref{KKcouplings}), in the
limit $c_{f} \rightarrow 1/2$ give a very good approximation to the
$c_{f}=1/2$ case (within a few percent).

For $\alpha_{Q} k \sim \alpha_{t} k$ of order a few, the decoupling of
the higher KK modes is very efficient and the contribution from the
first KK level is an excellent approximation to the full tower.  In
addition, for $c_{Q} = 1/2$, $c_{t} = 0$ and $\alpha_{Q,t} k \sim$~a
few, the diagram with $t_R$ zero modes dominates over that with $t_R$
KK modes.  Thus, we may approximate the complete fermionic
contribution to $\Delta T$ as,
\beq
\label{deltaTtop2}
\Delta T_t  \approx 
\left(\frac{\lambda_{10}}{\lambda_t} \right)^{2}
\left(\frac{m_t}{m^{(1)}_{t_L}}\right)^2 
\left[ \frac{N_c}{16 \pi s^2 c^2} \left(\frac{m_t}{m_Z}\right)^2\right]
\left\{ \frac{4}{3} \left(\frac{\lambda_{10}}
{\lambda_t} \right)^{2}
+ 4 \left[ 2 \log \left( \frac{m^{(1)}_{t_L}}{m_t}\right) - \frac{3}{2} \right]
\right\}
~,
\eeq
where the term in square brackets is the SM top contribution, which is
of order one.  The terms in the curly brackets are the expressions for
graphs containing two $t_L^{(1)}$ and one $t_L^{(1)}$ lines, 
respectively\footnote{The graph with one $t_L^{(1)}$ nominally contains
an IR divergence in the mass insertion approximation.  We have dealt
with this subtlety by resumming all insertions of the zero mode mass.}.
This expression 
is a good approximation to the entire KK fermionic contribution for
$\alpha k \gsim 3$.  For smaller $\alpha$, there are relevant
contributions from the $t_R$ KK modes.
For the choice $\alpha_{Q} k = \alpha_{t} k = 5$, $k
\LL = 30$, we find $\lambda_{t_{L}^{(1)},t_{R}^{(0)}}=2.37 \, \lambda_t$
and $m^{(1)}_{t_L} = 0.67 \, \tilde{k}$.  Taking, for example,
$\tilde{k} = 5~\textrm{TeV}$, results in $\Delta T \sim 0.25$.  We note that
the relevant coupling at the second KK level is
$\lambda_{t_{L}^{(2)},t_{R}^{(0)}} = 0.42 \, \lambda_t$, while
$m^{(2)}_{t_L} = 3.94 \, \tilde{k}$, from which it is easy to check
that its effect is completely negligible.  We conclude that the
fermion localized kinetic terms are very efficient in suppressing
these loop contributions to $T$, even for $c_{Q}=1/2$.

\subsection{KK Gauge Boson Contributions}  
\label{sec:gauge}  

\begin{figure}[t]  
\centerline{\includegraphics[width=0.8\textwidth]{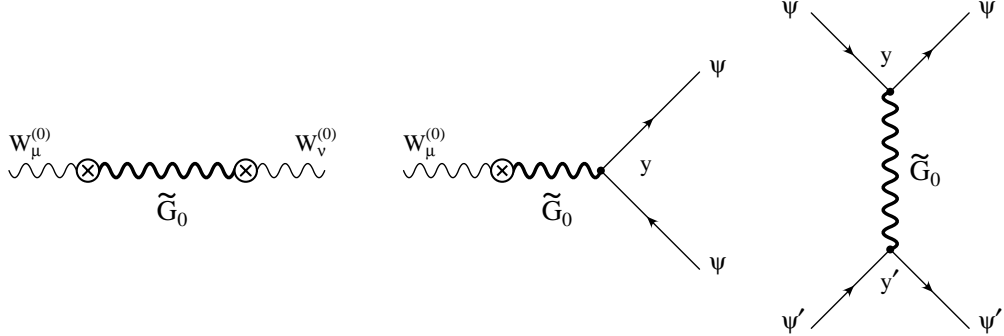}}
\caption{Tree-level contributions to the weak gauge boson masses, the
gauge-fermion couplings, and four fermion operators, arising from the
KK gauge bosons.  $\GG_{0}$ stands for the KK propagator at zero
momentum, and the crosses represent insertions of the Higgs VEV
squared
(at $y=L$) whereas the small filled circles are insertions of the bulk
$W$-$f$-$\overline{f}$ vertex, integrated over $y$.}
\label{fig:gauge}  
\end{figure}

Integrating out the KK gauge bosons leads to important tree-level
corrections to the weak gauge boson masses as well as to the couplings
among the gauge fields and the quarks and leptons.  These corrections
arise from the fermion couplings to the KK gauge bosons and as a
result of the mixing of the zero-mode weak gauge bosons with their KK
modes, induced by the presence of the localized Higgs fields, as
indicated in Fig.~\ref{fig:gauge}.

Such effects can be efficiently handled with the aid of the propagator
for the massive KK gauge fields, as explained in Ref.~\cite{CDPTW}.
The KK summation can be done automatically by working in mixed
position and 4D momentum space.  Denoting this propagator by
$\GG_{p}(y,y')$, where $p$ is the 4D momentum, the dominant low-energy
corrections are all determined by the KK gauge propagator evaluated at
zero momentum and the fermion zero-mode wavefunctions.  In detail, the
leading corrections may be computed in terms of $\GG_0^3(\LL,\LL)$,
$\GG_0^B(\LL,\LL)$, and the quantities
\beqa
\label{convolutions}
G_f^i &\equiv& \int_0^\LL dy \GG_0^i(\LL,y) |f^{(0)}(y)|^2 
\; \left(1 + 2 \alpha_{f} \delta(y-L) \right)
\nonumber \\
G_{ff}^i &\equiv& \int_0^\LL dy dy' |f^{(0)}(y)|^2 \GG_0^i(y,y') 
|f^{(0)}(y')|^2~
\; \left(1 + 2 \alpha_{f} \delta(y-L) \right)
\; \left(1 + 2 \alpha_{f} \delta(y'-L) \right)
,
\eeqa
where the superscript $i= 3,B$ in the above quantities refer to the
$W^3$ and $B$ gauge bosons of $SU(2) \times U(1)$, respectively, and
$f^{(0)}(y)$ is the appropriate fermion zero-mode wavefunction, given
in Eq.~(\ref{zeromodefermion}).  The terms proportional to
$\delta(y-L)$ represent the effects induced by the presence of the
gauge-covariant brane kinetic terms of the fermions.

Observe that while the quantities $G_f$ and $G_{ff}$ serve to
determine the corrections to the effective couplings of the zero-mode
fermions to the weak gauge bosons and the induced four fermion
operators, respectively, the corrections to the gauge boson masses are
just a function of $\GG_0(L,L)$.  For instance, the $Z$ and $W$ masses
are given by
\beq
\label{Zmass}
m_Z^2 = \frac{e^2 \vv^2}{2 s^2 c^2} \left\{1 + \frac{\vv^2}{2} 
[\GG^3_0(\LL, \LL) + \GG^B_0(\LL, \LL)] + {\cal{O}}(\frac{v^4}{k^4}) \right\}~.
\eeq
\beq
\label{Wmass}
m_W^2 = \frac{e^2 \vv^2}{2 s^2} \left\{1 + \frac{\vv^2}{2} 
\GG^3_0(\LL, \LL) + {\cal{O}}(\frac{v^4}{k^4})
\right\}~.
\eeq
In the above $\vv = v e^{-kL} \simeq 174~\textrm{GeV}$ is the Higgs
field vacuum expectation value, and $s$ and $c$ represent the sine and
the cosine of the tree-level weak mixing angle, $c =
g/\sqrt{g^2+g'^2}$ and $s = g'/\sqrt{g^2+g'^2}$.

Finally, the Fermi constant is given by
\beqa
\label{G_F}
2 \sqrt{2} G_F = \frac{1}{\tilde{v}^2} \left[ 1 + \frac{\tilde{v}^2}{2} 
\left( 2 G_f^{3} -  \GG_0^{3}(L,L) - G_{ff}^{3} \right) 
+ {\cal{O}}(\frac{v^4}{k^4}) \right]~,
\eeqa
where the $G_{ff}$ term represents non-oblique corrections. 

In the cases we are going to analyze, with universal $c_{f}$ and
$\alpha_{f}$ parameters, the only relevant non-oblique corrections to
$m_W$ and the $Z$-pole observables come indirectly through the Fermi
constant $G_F$.  In this case, one can define effective precision
electroweak parameters which determine $m_W$, as well as all Z-pole
observables.  Following Refs.~\cite{Carena:2002dz,CDPTW}, and
considering the expression of $G_F$, Eq.~(\ref{G_F}), it is possible
to define these effective parameters $S_{\rm{eff}}$, $T_{\rm{eff}}$
and $U_{\rm{eff}}$, which are given by,
\beqa
\label{STUeffective}
\alpha_{\rm em} S_{\rm{eff}} &\approx& 2 \vv^2 
[s^2 G_f^3 + c^2 G_f^B] + {\cal{O}}(\frac{v^4}{k^4})~, \nonumber \\
\alpha_{\rm em} 
T_{\rm{eff}} &\approx& \frac{\vv^2}{2}  [2 G_f^B - \GG_0^B(\LL,\LL)
+ G^3_{ff}] + {\cal{O}}(\frac{v^4}{k^4})~,  \\
\alpha_{\rm em}  U_{\rm{eff}} &\approx& 
- 2 s^2 \vv^2 G^3_{ff} + {\cal{O}}(\frac{v^4}{k^4})~, \nonumber
\eeqa

It is straightforward to find the explicit expression for $G_f$,
$G_{ff}$ and $\GG_0(\LL,\LL)$ in terms of the fundamental parameters
of the theory.  In the following, for simplicity, we ignore the UV
brane kinetic terms.  We comment on their effects below.
In this case, one finds 
\beq
\label{GLL}
 \GG_0(\LL, \LL) = - \frac{e^{2k\LL} g^2 }{k^2}
\frac{2 k^2 \LL^2 - 2 k\LL  + 1}
{4 k (\LL + r_{IR})} ~.
\eeq
For further details of the calculation of the KK gauge propagator,
$\GG_{p}(y,y')$, and its use to obtain the low-energy effective
theory, refer to \cite{Carena:2002dz,CDPTW}.

The expressions for $G_f$ and $G_{ff}$ depend on the parameters $c_f$
and $\alpha_{f}$ through the fermion zero-mode wavefunctions,
Eq.~(\ref{zeromodefermion}).  The exact analytic expressions can be
obtained in a straightforward manner, although the general results
have a somewhat complicated dependence on $c_f$.  However, in the case
of interest here, where $c_{f} \gsim 1/2$, the expressions simplify
considerably, up to exponentially small terms. We find

$\bullet$ For $c_f - \frac{1}{2} > 1/2k\LL$: 
\beqa
\label{GfUVfermions}
G_f &=& \frac{e^{2k\LL} g^2 }{k^2} \frac{k\LL - 1 - k r_{IR} +
2 k^2 r_{IR} \LL}{4 k (\LL + r_{IR})}~, \\
\label{GffUVfermions}
G_{ff} &=& - \frac{e^{2k\LL} g^2 }{k^2} \frac{2 k^2 r_{IR}^2 + 2 k r_{IR} +
1}{4 k (\LL + r_{IR})} ~,
\eeqa
where $g$ is the (zeroth order) zero-mode gauge coupling.  Note that
the results in Eqs.~(\ref{GfUVfermions}) and (\ref{GffUVfermions}) are
independent of $c_f$ and $\alpha_{f}$.

$\bullet$ For $c_f = \frac{1}{2}$: 
\beqa
\label{GfFlatfermions}
G_f &=& \frac{e^{2k\LL} g^2 }{k} 
\frac{(2 k^2 \LL^2-2 k \LL+1) (r_{IR} 
- \alpha_{f})}{4 k^2 (\LL + r_{IR}) (\LL + \alpha_{f})}~,  \\
\label{GffFlatfermions}
G_{ff} &=& - e^{2k\LL} g^2 \frac{
(2 k^2 \LL^2 - 2 k \LL +1) (r_{IR}-\alpha_f)^2}{4 k^3 (\LL + r_{IR}) (\LL+\alpha_{f})^2}~.
\eeqa

Note that, in this case, $G_{f}$ may have either sign depending on the
relative size of the gauge and fermion kinetic terms, $r_{IR}$ and
$\alpha_{f}$.  Also note that Eqs.~(\ref{GfFlatfermions}) and
(\ref{GffFlatfermions}) vanish when $r_{IR} = \alpha_{f}$.  This is a
consequence of the fact that, in this case, the fermion (see
Eq.~(\ref{normalization}) for $f_{L}^{n}$) and gauge orthogonality
conditions are identical, and the fact that for $c_{f}=1/2$ the
fermion zero-mode wavefunction is flat and therefore proportional to
the gauge zero mode wavefunction.  Thus, the coupling of the zero-mode
fermions to the higher KK gauge modes vanishes identically in this
limit.  This same fact was recently observed in warped extra-dimensional
Higgsless models~\cite{Cacciapaglia:2004rb}.

The precision electroweak observables depend on the relative size of
the parameters $G_f$, $G_{ff}$ and $\GG_0(L,L)$.  In the limit of
large values of $\alpha_{f} \gg L$, and for $c_f \leq 1/2$, the values of
$G_f$, $G_{ff}$ tend to $\GG_0(L,L)$.  This result coincides with the
one associated with fermions localized in the infrared brane.  What
happens in this case is that the physics is governed by the effects
induced by the four dimensional brane kinetic terms, and propagation
in the bulk becomes unimportant.  The case of fermions localized in
the infrared brane was already analyzed in 
\cite{Csaki:2002gy,Davoudiasl:2002ua,Carena:2002dz}.

Assuming that all quark and lepton bulk mass parameters, other than
the right-handed top quark one, take values $c_f \simeq 1/2$ and that
there is a common brane kinetic term coefficient $\alpha_{f} \equiv
\alpha$, simple analytical expressions for $T_{\rm eff}$, $S_{\rm
eff}$ and $U_{\rm eff}$ may be obtained:
\begin{equation}
T_{\rm eff} \simeq \frac{\pi}{c^2}
\left(\frac{\tilde{v}}
{\tilde{k}}\right)^2
\left[\frac{k(L + 2 r_{IR} - \alpha)}{(1 + r_{IR}/L)
(1 + \alpha/L)}\right] - \frac{ U_{\rm eff}}{ 4 s^2} ,
\label{teff}
\end{equation}
\begin{equation}
S_{\rm eff} \simeq 8 \pi  \left(\frac{\tilde{v}}
{\tilde{k}}\right)^2
\left[\frac{k(r_{IR} - \alpha)}{(1 + r_{IR}/L)
(1 + \alpha/L)}\right],
\label{seff}
\end{equation}
\begin{equation}
U_{\rm eff} \simeq \frac{S_{\rm eff}}{2} \left[ \frac{r_{IR}/L - \alpha/L 
}
{1 + \alpha/L} \right] ,
\label{ueff}
\end{equation}
where, taking into account that $k L \simeq 30$, we have ignored terms
of order $1/(kL)$.  For moderate values of $k \alpha , kr_{IR} \ll
k L$, the non-oblique corrections to the precision electroweak
observables become small and, in particular, $U_{\rm eff}$, Eq.~(\ref{ueff}), 
become much smaller than $S_{\rm eff}$ and can be safely
neglected in the description of the new physics corrections to the
precision electroweak observables.

It is also useful to have an analytical approximation for the top-quark
KK mode contribution to the parameter $T_{\rm eff}$, valid in the limits
in which $m_{KK}/ \widetilde{k} \lsim 1$ 
and $\alpha k \gsim 1$.  In order to do this
we computed the mass of the first KK mode of the left-handed top
quark, for $c_f = 1/2$, as a function of the brane kinetic term
coefficient $\alpha$ (The same expression is valid for the mass of the
first gauge field KK mass as a function of $r_{IR}$).
\begin{equation}
\label{KKmass}
m^{(1)}_{t_L} 
\simeq k  e^{-k L} \sqrt{8 \frac{(1 + \alpha/L)}{(1 + 4 k \alpha)}}
\end{equation}
while
\begin{equation}
\frac{\lambda_{t_{L}^{(1)},t_{R}^{(0)}}}{\lambda_t} \simeq
2 \; \sqrt{\frac{kL}{1 + 4 k \alpha}}.
\label{lambdat1t0}
\end{equation}
Therefore, Eq.~(\ref{deltaTtop2}) reduces to
\begin{equation}
\Delta T_{t} \approx 
\left( \frac{kL}{1+\alpha/L} \right) \frac{m_t^2}{\widetilde{k}^2}
\left[ \frac{N_c}{16 \pi s^2 c^2} \left(\frac{m_t}{m_Z}\right)^2\right]
\left\{ \frac{8 k L}{3 (1+4 k \alpha)} + 2 
\log \left( \frac{8 \tilde{k}^2 (1+\alpha/L)}{m_t^2 (1+4 k \alpha)} \right)
-3 \right\} ~,
\label{T-top}
\end{equation}
where, as before, the term in the square brackets is the SM contribution to
$\Delta T$ from top, and is approximately equal to 1.2.

Up to now, we have neglected the effect of UV brane localized kinetic terms.
We now briefly comment on their effects. Gauge and fermion localized
UV kinetic terms have a mild impact on the spectrum. The effect of the UV
terms in the KK masses amounts to replace $\LL$ by $\LL + \alpha_{UV}$
in Eq.~(\ref{KKmass}) (or equivalently by $\LL + r_{UV}$ for the gauge boson
KK masses). The main effect of UV kinetic terms is to shift the contribution
of gauge and fermion KK modes to precision electroweak observables
by quantities of order $r_{UV}/\LL$ ( or $\alpha_{UV}/\LL$). Hence,
provided they are smaller than $\LL$, the inclusion of $r_{UV}$ and 
$\alpha_{UV}$ do not  change the values of $\tilde{k}$ and the KK gauge
boson and fermion masses  consistent with  experimental data in a 
significant way. 

For completeness, we present the expressions of $\GG_0(\LL,\LL)$ 
and $G_f$ for non-vanishing values of the UV kinetic terms.
Keeping only dominant terms, for $k r_{IR}$ of order a few,
$\GG_0(\LL,\LL)$ reads~\cite{CDPTW},  
\beq
\label{GLLruv}
 \GG_0(\LL, \LL) \simeq - \frac{g^2 }{2 \tilde{k}^2} \;
\frac{ k^2 (\LL + r_{UV})^2}
{k (\LL + r_{IR} + r_{UV})} ~.
\eeq
As anticipated, the comparison of this expression  with the one presented in 
Eq.~(\ref{GLL}) shows corrections of order $r_{UV}/\LL$.
Note that $\GG_0(\LL \LL)$ is independent of the details associated with the 
fermion sector, and that it only affects the effective $T$ parameter.
On the other hand, 
the size of $G_f$ and $G_{ff}$ are
controlled by the IR parameters, $r_{IR}$ and $\alpha_{f}$, and
receive corrections from
$r_{UV}$ (gauge) and $\alpha_{UV}$ (fermion) kinetic terms 
of order $r_{UV} / \LL$ and $\alpha_{UV}/\LL$, respectively,
but with a less straightforward
dependence than the $\GG_0(\LL,\LL)$ one. For instance,
in the case $c_f = 1/2$ and for $k r_{IR}, k \alpha_{f}$ of order a few, 
the dominant contribution 
to $G_f$ reads
\beqa
\label{Gfruv}
G_f & \simeq & \frac{g^2 }{\tilde{k}^2} \;
\frac{k (L+ r_{UV})[r_{IR}(L + \alpha_{UV}) - 
\alpha_f (L+ r_{UV})]}
{2 (\LL + r_{IR} + r_{UV}) (\LL + \alpha_{f} + \alpha_{UV})}~,  
\eeqa
Notice that the orthogonality condition, that ensures the
cancellation of $G_f$, is fulfilled for
$r_{IR} = \alpha_{f}$ and $r_{UV} = \alpha_{UV}$.

Finally, let us mention that for $k \alpha_f$ of order a few
the expression of 
$\lambda_{t_{L}^{(1)},t_{R}^{(0)}}/\lambda_t$ for non-vanishing
values of $\alpha_{UV}$ may be obtained by changing $L$
by $L + \alpha_{UV}$ in Eq.~(\ref{lambdat1t0}). Moreover,
the contributions of $G_{ff}$ to $T$ and 
$U$ remain very small provided $k\, r_{IR}$ and $k \alpha_{IR}$ are 
less than order a few, 
even if $r_{UV}$, $\alpha_{UV}$ are as large as order $\LL$.

Sizable UV localized gauge kinetic terms appear, for instance, 
in the unification scenario analyzed in Ref.~\cite{CDPTW}, 
where $r_{IR}^B + r_{UV}^B \simeq - (r_{IR}^3 +r_{UV}^3) \simeq
L/3$.
In this particular case, for moderate values of the IR kinetic 
terms, $k r_{IR} \lsim 2$ (taking $r_{IR}^B = r_{IR}^3 = r_{IR}$),   
and $\alpha_{UV} = 0$, one obtains corrections
of about 30 percent to the gauge boson contributions to the 
$T$ parameter, while the correction to the
$S$ parameter are smaller than 10 percent.
One would then find corrections of about 10--15 percent for the 
values of $\tilde{k}$ consistent with the electroweak observables. 
Since the relation between the KK masses and $\tilde{k}$ is quite
insensitive to the UV localized terms, this translates directly into a
10--15 percent correction to the KK masses. 
This should be compared with the 
effects induced by IR kinetic terms, that modify the relation between the 
lightest KK masses and $\tilde{k}$ in a much more crucial way, and 
control the value of the effective $S$ parameter, as well as the 
top-quark KK mode contributions to the $T$ parameter.
Therefore, for simplicity, in the following section we shall
restrict ourselves to the case of vanishing values of 
the UV kinetic terms.

\subsection{Electroweak Fit}  
\label{sec:ew}  

In this section, we will consider the case in which the values of
$\alpha \, k$ and $r_{IR} k$ are of order of a few, and hence
for vanishing values of the UV kinetic terms
Eqs.~(\ref{deltaTtop2}), (\ref{teff}), (\ref{seff}) and (\ref{ueff})
provide a good description to the dominant fermion and gauge boson
contributions to the precision electroweak data.  In this case, the
model under consideration falls under the general class of theories in
which there are only small corrections to the parameter $U$, while the
corrections to $S$ and $T$ are sizable and of the order of the
corrections associated with a heavy Higgs boson.  One can therefore
extract the allowed values of $S$ and $T$, by making a fit to the
electroweak precision data under the assumption that all new physics
contributions can be parametrized by these two parameters.

While making a fit to the electroweak data, one must choose a
reference value for the Higgs boson mass, $m_{H_{\rm ref}}$, for which
the SM gives $S = T = 0$. One then obtains a countour in the $S$, $T$
plane indicating the allowed new physics contributions to the $S$ and
$T$ parameters for that particular value of that Higgs mass. Had
one chosen a different reference value for the Higgs mass, $m_H$, 
the allowed new physics contribution to $S$ and $T$  would 
be shifted by an amount equal (but of opposite sign) to the contribution
to $S$ and $T$ obtained by the change of the Higgs mass from 
$m_{H_{\rm ref}}$ to $m_H$. This Higgs boson contribution to $S$ and $T$
is given by
\begin{eqnarray}
S_H = \frac{1}{12 \pi} \ln \left(\frac{m_H^2}{m_{H_{\rm ref}}^2}\right)
\, ,
\nonumber\\
T_H = -\frac{3}{16 \pi c^2} \ln \left(\frac{m_H^2}{m_{H_{\rm ref}}^2}\right)
\, .
\end{eqnarray}

\begin{figure}[htb]
\centerline{
        \includegraphics[width=0.9\textwidth]{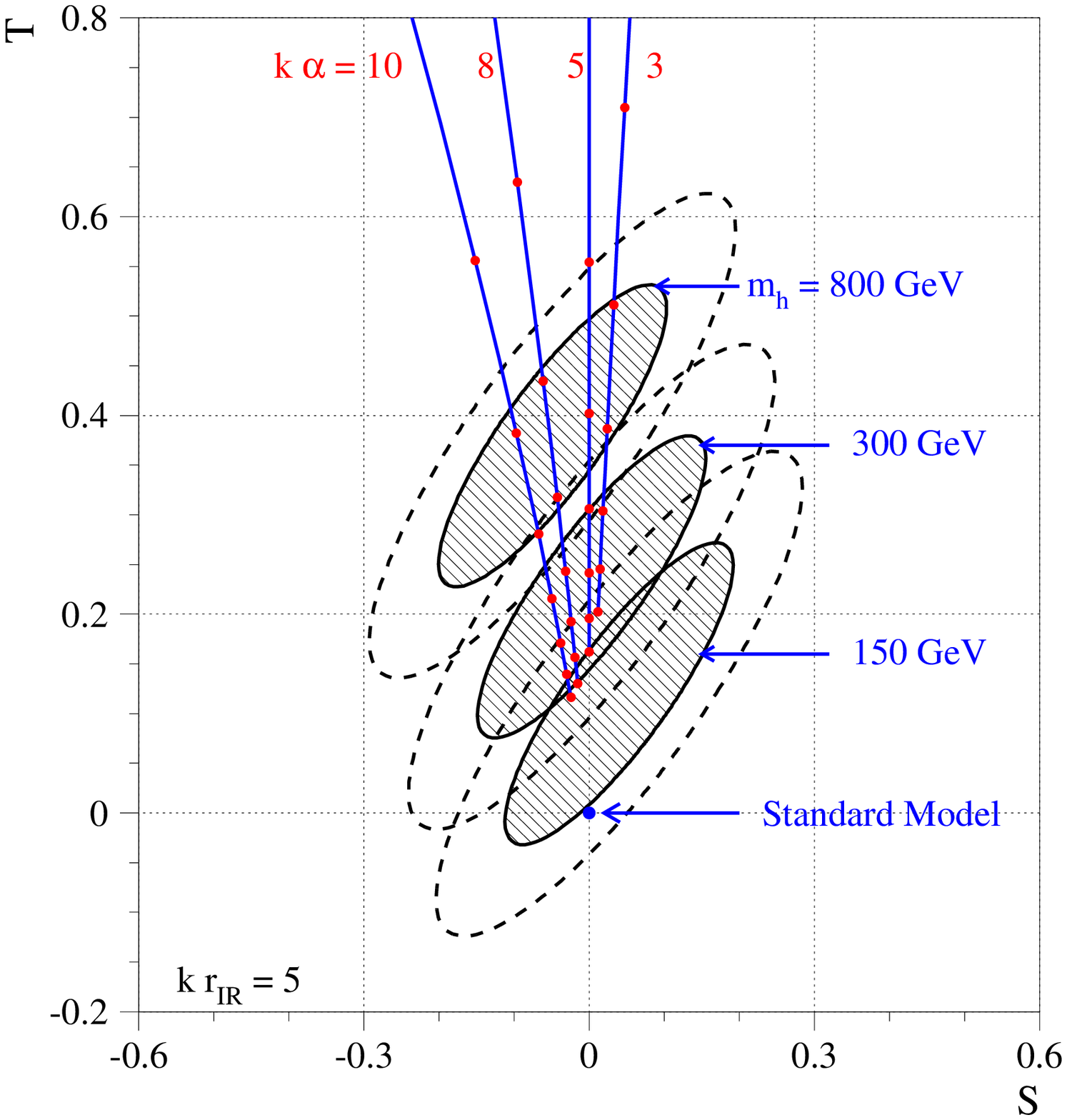}}
\vspace*{-2cm}
\caption{95\% and 68\% allowed regions in the $S$-$T$ plane, for 
$k r_{IR} = 5$, $r_{UV}^3 = r_{UV}^B = 0$, and 
different values of the Higgs boson mass.  Also shown in the Figure is
the KK mode contributions to the $S$ and $T$ parameters for different
values of $\alpha$ and $\tilde{k}$, starting (at the lower end
of each fixed $\alpha$ line) with $\tilde{k}=10$~TeV and
decreasing in steps of 1 TeV for each dot.}
\label{tblam}        
\end{figure}

We are interested in setting constraints on the masses of the fermion
and gauge boson KK excitations, for arbitrary values of the Higgs
boson mass.  The LEP electroweak working group has recently extracted
the allowed values of the $S$ and $T$ parameters coming from a fit to
the electroweak precision data \cite{EWWG}.  For a reference value of
the Higgs boson mass $m_{H_{\rm ref}} =$ 150~GeV, they obtained
\begin{eqnarray}
S \simeq 0.04 \pm 0.10 \, ,
\nonumber\\
T \simeq 0.12 \pm 0.10 \, ,
\end{eqnarray}
with $85\%$ correlation between the two parameters.  Based on this
information, in Fig.~\ref{tblam} we obtain the $95\%$ confidence level
allowed region for the $S$ and $T$ parameters for three different
values of the Higgs bosons mass $m_{H} = 115, 300$ and 800
GeV, respectively.  Also shown in the Figure are the KK mode
contributions to the $S$ and $T$ parameters for different values of
$\alpha$ and $\tilde{k}$, for a value of the gauge field brane kinetic
term $k r_{IR} \simeq 5$.  We see that generically, a heavier Higgs
boson mass allows for lower values of $\tilde{k}$, due to compensation
between contributions to $T$ from the Higgs and the extra dimensional
contributions.  Values of $\tilde{k}$ as low as 4.5 TeV are consistent
with experimental data. Note that for $\alpha k \sim r_{IR} k \sim 5$,
the mass of the first KK modes are about $2/3 \times \tilde{k}$,
whereas for $\alpha k \sim 10$ the fermion first KK mode mass is
approximately $1/2 \times \tilde{k}$. Hence for these particular
values of the IR kinetic terms, the lightest KK gauge
boson and fermions masses may be as low as about 3~TeV and 2.3~TeV, 
respectively.


\section{Conclusions}  
\label{sec:conclusion}  

Extra dimensional models provide an alternative solution to the gauge
hierarchy problem.  Among the different realizations of this idea, the
Randall Sundrum model is perhaps the most attractive one.  In
particular the Randall Sundrum model with fermions and gauge bosons
propagating in the bulk allows to address the question of unification
of couplings and sets the framework for a possible understanding of
flavor coming from the localization of the fermions in the bulk of the
extra dimensional space.

In this article we have studied the impact of localized brane kinetic
terms for the fermions in this scenario.  The infrared brane kinetic
terms repell the wavefunction of the heavy KK modes from the infrared
brane where the Higgs field is localized and allows to solve the
strong coupling problem of the top Yukawa sector and to minimize
potentially dangerous flavor-violating effects.  It is interesting to
see that despite its underlying non-renormalizability, the extra
dimensional theory already contains in itself a mechanism to suppress
power-law corrections to brane couplings.

In the same spirit, a fermion brane kinetic term further renders the
potential quadratically divergent contributions to the $T$ parameter
finite and reduces the impact of the extra dimensional effects on the
precision electroweak parameters.  This allows all of the left-handed
fermions to have bulk masses with $c \gsim 1/2$, and allows one to
realize the attractive scenario in which the SM flavor hierarchies are
(at least in part) generated by extra-dimensional geometry.  Previous
attempts have had larger corrections to the $Z$ coupling to bottom
quarks, in contradiction with high precision measurements.  In the
end, KK masses as low as a few TeV are permitted, which could be
discovered at the LHC.


~\\  
{\Large \bf Acknowledgements}\\  
~\\  
Work at ANL is supported in part by the US DOE, Div.\  
of HEP, Contract W-31-109-ENG-38.  Fermilab is operated by  
Universities Research Association Inc.  under contract no.  
DE-AC02-76CH02000   
with the DOE. A.~D.~ is supported by NSF Grants  
P420D3620414350 and P420D3620434350   
and also wants to thank the Theory  
Division of Fermilab and IFT Madrid
for the kind invitation.

\end{document}